\documentclass[aps,showpacs,prb,twocolumn]{revtex4}

\usepackage{graphicx}
\usepackage{dcolumn}
\usepackage{bm}
\usepackage{color}

\begin{document}
\title{Low Temperature Transport in Undoped Mesoscopic Structures}
\author{S. Sarkozy}
\altaffiliation[stephen.sarkozy@ngc.com (sjs95@cam.ac.uk) on leave from ]{Northrop Grumman Space Technology, One Space Park, Redondo Beach, California, 90278}
\author{K. Das Gupta}
\author{C. Siegert}
\author{A. Ghosh}
\altaffiliation[Present address ]{Department of Physics, Indian Institute of Science, Bangalore 560 012, India}
\author{M. Pepper}
\author{I. Farrer}
\author{H. E. Beere}
\author{D. A. Ritchie}
\author{G. A. C. Jones}
\affiliation{Cavendish Laboratory, J.J. Thomson Avenue, Cambridge CB3 0HE, United Kingdom}
\date{\today}

\begin{abstract}
{Using high quality undoped GaAs/AlGaAs heterostructures with optically patterned insulation between two layers of gates, it is possible to investigate very low density mesoscopic regions where the number of impurities is well quantified. Signature appearances of the scattering length scale arise in confined two dimensional regions, where the zero-bias anomaly (ZBA) is also observed. These results explicitly outline the molecular beam epitaxy growth parameters necessary to obtain ultra low density large two dimensional regions as well as clean reproducible mesoscopic devices.}
\end{abstract}

\maketitle
With potential applications to quantum computation as well as innate interest in fundamental physics such as localization and Quantum Hall effects, significant resources have been devoted to the study of electron-electron interactions\cite{Pepper:Localization,RevModPhys.70.1039}. Regardless of dimensionality, the relative importance of interactions increases with decreasing carrier concentration, providing substantial impetus for the study of low density systems. It is instructive to quantitatively analyze the requirements to ensure that phenomena observed are intrinsic to interactions in the two dimensional electron/hole gas (2DEG/2DHG), and not merely manifestations of impurity effects or disorder based localization. The cleanliness of the 2D gas at low densities is also crucial in studying pinch-off regions in one-dimensional (1D) channels and very low conductance characterization of zero-dimensional (0D) quantum dots, although this point is not often emphasized. The theory of scattering mechanisms in 2DEGs is very mature\cite{2DScatteringRevModPhys.54.437,GoldPhysRevB.38.10798,AndoIR}, providing a concrete basis for improvements necessary to successfully probe regimes of strong interaction. In particular, the proximity of ionized dopant impurities (modulation or $\delta$-doping, or even a doped cap) significantly lowers mobility at low carrier densities, as well as contributes to Random Telegraph Signal noise. Overcoming mobility degradation in bulk 2D studies is possible by offsetting the dopants from the heterointerface by an extremely thick spacer layer\cite{PhysRevLett.90.056806}, but an offset of more than $100 - 200nm$ becomes untenable when defining certain features of mesoscopic structures, such as narrow quantum point contacts ($\sim500nm$) or quantum dots ($\sim50nm$). The alternative method of doping a quantum well from underneath only, creating an inverted 2D gas, carries the cost of interface roughness scattering from inverted interfaces, and does not yield the best mobilities.
\begin{figure}
    \includegraphics[width=\columnwidth]{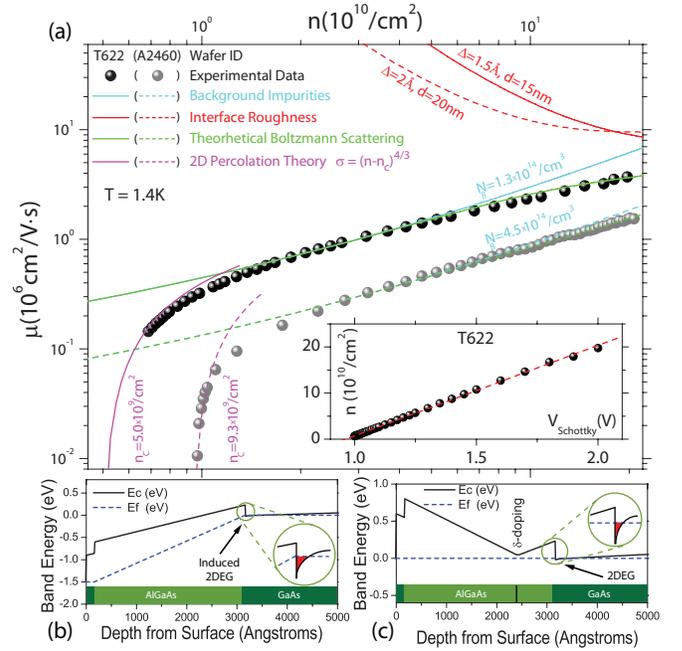}
        \caption{
             (a) Boltzmann Transport modeling of experimental $\mu-n$ data from two wafers (T622, A2460) assuming background impurities and interface roughness as the predominant scattering mechanisms in the linear screening regime, and percolation type behavior for the lowest carrier densities. (inset) Carrier density is linearly dependent on gate voltage. (b) In undoped heterostructures, electrons are induced via a positive bias on a top gate. The Fermi level is not constant through the system. (c) In n-type $\delta$-doped heterostructures carriers arise from donors, and the conduction band minima dips at the donor layer.
        }\label{Fig1}
\end{figure}

\begin{figure}
    \includegraphics[width=\columnwidth]{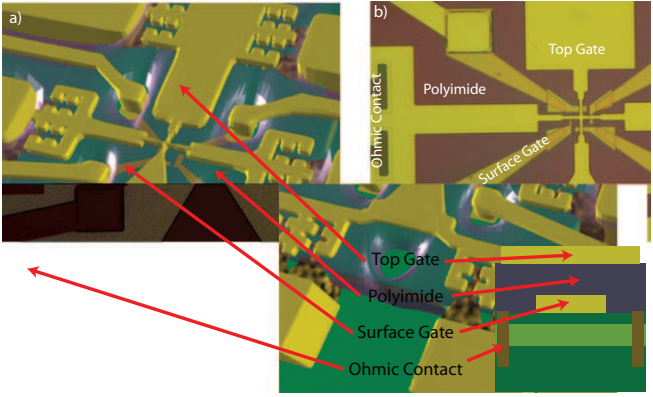}
        \caption{
            (Color online) Schematic (a) and optical image (b) of a device set, demonstrating the multiple gate layer process used to fabricate lower dimensional mesoscopic structures.
        }\label{Fig2}
\end{figure}
This Letter investigates the consideration necessary for integrating sharp lithographic features with a fully undoped structure, using a multi-level (top and Schottky) gating technique\cite{harrell:2328}. We also quantitatively estimate the effects of background impurities using Boltzmann Transport and classical probability. The fundamental difference between induced\cite{kane:2132,willett:242107,lu:112113} and doped systems is illustrated by the band structures shown in Fig.\hspace{.1cm}\ref{Fig1}(b) and (c). Unlike in $\delta$-doped structures, in induced devices the Fermi level is not constant and the 2D gas only forms when a threshold inducing voltage in reached - electrons/holes are pulled from the n/p type ohmic contacts by a biased top gate in a principle analogous to that of enhancement mode MOSFETs. The high quality of the the 2DEG can be judged from the fact that (for T622) at a carrier density $n=3\times 10^{10}cm^{-2}$ the mobility $\mu>10^6cm^2V^{-1}s^{-1}$, ($k_Fl=125$) and at $n=10^{11}cm^{-2}$ $\mu=3\times10^6cm^2V^{-1}s^{-1}$ ($k_Fl>1000$)\cite{Pfeiffer:Growth}. The number of ionized background impurities ($N_B$) at the heterointerface can be calculated from the experimentally observed variation of mobility with carrier density. Assuming that {\it all} the scattering comes from the presence of these impurities and interface roughness,(which the fit in Fig.\hspace{.1cm}\ref{Fig1}(a) justifies)  Mattheisen's Rule gives the inverse scattering time \mbox{$\frac{1}{\tau}$} as the sum of
\begin{equation}
\mbox{$\frac{1}{\tau_B}=\frac{N_B}{2{(2k_{F})}^{3}}\frac{m^*}{\pi\hbar^3}(\frac{e^2}{2\epsilon_0\epsilon_r})^2\int_{0}^{\pi} d\theta\frac{1-\cos\theta}{\epsilon(q)^2 \sin^3\theta/2}$}
\end{equation}
 and
 \begin{equation}
 \mbox{$\frac{1}{\tau_{IR}}=\frac{1}{{(2k_{F})}^{2}}\frac{2m^*}{\hbar^3}(d\Delta)^2\int_{0}^{\pi} d\theta \frac{q^2\Gamma(q)^2}{exp(\frac{d^2q^2}{4})\epsilon(q)^{2}}$}
 \end{equation}
 with commonly defined\cite{AndoIR} statistical interface roughness parameters $\Delta$ the mean square roughness height and $d$ a Gaussian correlation. The dielectric function $\epsilon(q)$ at a certain scattering wave vector $q=2k_F\sin\theta/2$ can be evaluated using the Fang-Howard form factor $F(q)$ and the polarizability $\Pi(q)$ of the electron gas as
 \begin{equation}
 \mbox{$\epsilon(q)=1 + \frac{F(q)\Pi(q)}{2\epsilon_0\epsilon_r}\frac{e^2}{q}$}
 \end{equation}
 with the dielectric constant of GaAs $\epsilon_r=12.8$. Obtaining from this $N_B \approx 1.3\times10^{14}cm^{-3}$ and using a wavefunction of calculated width $\lambda \leq 20nm$ gives an average distance between scattering centers $D \approx 600nm$. The importance of the background impurity concentration is not limited to restricting the mobility at low carrier densities, but also plays a role in determining the localization threshold or the onset of inhomogeneity, where the assumption of linear screening in the 2DEG breaks down. As can be seen in Fig.\hspace{.1cm}\ref{Fig1}(a), by reducing the background concentration from $4.5\times10^{14}cm^{-3}$ in A2460 to $1.3\times10^{14}cm^{-3}$ in T622 shifted the critical density from $\sim 9.3\times10^{9}cm^{-2}$ to $\sim 4.6\times10^{9}cm^{-2}$. Recent theoretical requirements for reaching mobilities of $10^8cm^2V^{-1}s^{-1}$ have put great emphasis on the need to reduce scattering from dopants\cite{hwang:235437}. Although this is indeed vital, at the densities which many Fractional Quantum Hall Effect\cite{Pfeiffer:Growth} and Microwave Induced Resistance Oscillations\cite{ManiNature} experiments are carried out ($2-4\times10^{11}cm^{-2}$), scattering due to interface roughness must also be addressed. Even with high quality interfaces buffered by a thick superlattice, in order to fit our experimental data a roughness of nearly a quarter of a monolayer needed to be assumed, which limited the mobility to under $10^7cm^2V^{-1}s^{-1}$ at $n=3\times10^{11}cm^{-2}$ in the set of samples studied.

\begin{figure}
    \includegraphics[width=\columnwidth]{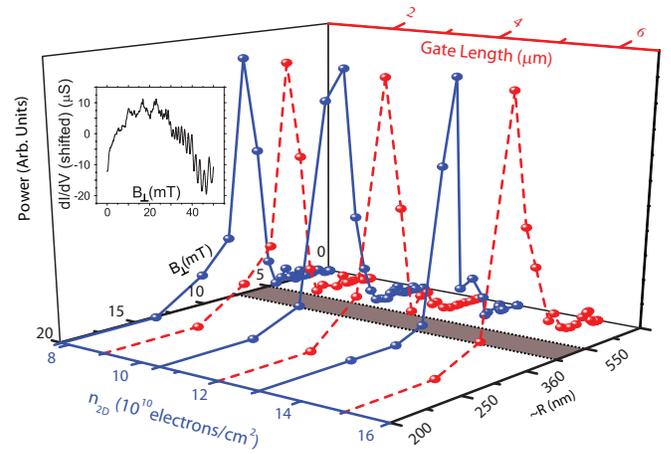}
        \caption{
            (Color online) Fourier Transform of normalized magnetoconductance traces for (dotted red) devices of varying gate lengths at $V_{TG} = 4V$ and (solid blue) a $2 \mu m \times 6 \mu m$ device at varying bulk $2D$ densities, correlated to incident electron velocity. The movement of the peaks is negligible, indicating a stable $R$. (inset) Typical shifted magnetoconductance trace showing the distinction between Aharonov-Bohm like and Shubnikov-de Haas oscillations.
        }\label{Fig3}
\end{figure}

Devices were fabricated from T622, an undoped Al$_{.33}$Ga$_{.67}$As/GaAs heterostructures with a 300$nm$ thick Al$_{.33}$Ga$_{.67}$As barrier grown upon a GaAs substrate. Fig.\hspace{.1cm}\ref{Fig2}(a) sketches the principle outlay of the devices. A mesa etch establishes isolation between different chip segments before recessed ohmic contacts (Ni/AuGe/Ni/Ti/Pt), patterned gold/black in the image, are deposited and annealed. The recessed nature of the contacts, by which they are deposited into etched pits in a self-aligned process, is crucial, as the required diffusion to contact the induced 2DEG is lateral, rather than the standard vertical diffusion in doped devices. Particular attention must be paid to the roughness of the contact edges in order to achieve high yield\cite{Sarkozy:Electrochem}. Non-magnetic surface gates (Ti/Au) are defined with electron beam lithography and deposited. An insulating layer of optically patterned polyimide (PI2737, HD Microsystems), depicted in the figure as a semi-transparent purple layer, provides separation from a top gate (Ti/Au) just overlapping the edge of the ohmic contacts. By restricting the overlap to the edge of the ohmic contact, the leakage (proportional to the overlap area) is minimized, while still maintaining the quality of the contact to the 2DEG (proportional to the perimeter).

Our objective is to study the Zero-Bias Anomaly (ZBA), an enhanced conductance at zero source drain bias, was recently observed in $\delta$-doped 1D and 2D mesoscopic systems\cite{Cronenwett:ZBA,Siegert:NaturePhysics,sfigakis:026807} and has attracted significant attention due to the Kondo-like similarities between the same effect in 0D\cite{Cronenwett07241998}. In addition to investigating the length scale of the background impurities in mesoscopic regions, study of the ZBA in a dopant-free system may shed light on its origin in situations without an obvious source of a bound state. A positive bias $V_{TG} > +3$ on the top gate induces electrons in the area underneath it, forming a well defined conducting path around the ohmic contacts under the top gate, but discontinuous under the surface gates unless $V_{SG}$ is positively biased. The density under the surface gates can be kept much lower than the bulk density. The width of the top gate region is $\sim6\mu m$ whereas the surface gates range from $2-8\mu m$. Measurements were performed in a two probe ac+dc setup with excitation voltage $V_{ex} < k_B T_{\rm base}$ in a dilution refrigerator with base temperature $T_{\rm base} \approx 30mK$ and at zero magnetic field unless explicitly stated. The electron temperature was approximated by temperature dependence saturation as $T_{\rm electron} \approx 60mK$.

\begin{figure}
    \includegraphics[width=\columnwidth]{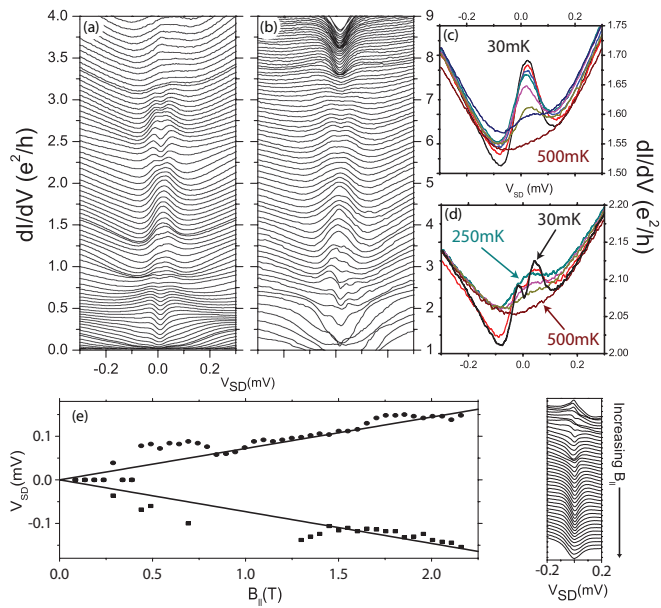}
            \caption{
            (Color online) left) ZBA trace for a $6 \times 6 \mu m^2$ device with $V_{TG} = 4V$. Clearly split peaks (type II ZBA) are seen. right) ZBA in a different device of similar size and top gate voltage, but from a different chip.  Here ZBA is seen up to $ \sim7e^2/h$, and at higher conductances, we see a significant dip in $G(V_{SD} = 0$).a) left) Monotonic temperature suppression for type I ZBA. right) Type II ZBA devolves into a single peak (cyan trace) by $T \sim 250mK$, before being suppressed like type I. c) Zeeman splitting of a type I ZBA, with peak positions plotted over parallel magnetic field.
        }\label{Fig4}
\end{figure}

Periodic oscillations in the magnetoconductance are observed in low strength perpendicular magnetic fields. These oscillations are distinct from Shubnikov-de Haas Oscillations, which are periodic instead in 1/B, and in fact Fig.\hspace{.1cm}\ref{Fig3}(a) shows a device where both oscillations can be seen. It has been postulated that this periodicity is related to the radius $R$ of a stable Aharonov-Bohm ring by the relationship\cite{SchusterPhysRevB.49.8510,WeissPhysRevLett.70.4118} $\Delta B_ \perp \approx h/ \pi eR^2$. Fourier Transforming confirms the periodicity of these oscillations, and yields a radius $R\approx500nm$, well in line with the calculated average distance between background impurities with which conduction electrons could interact. That this is indeed likely, and that the radius is not some other characteristic dependent on the region, can be seen in Fig.\hspace{.1cm}\ref{Fig3}(c), where $R$ is seen to be invariant under changes in bulk electron density (a factor of two) and device gate length (a factor of three).

Fig.\hspace{.1cm}\ref{Fig4}(a,b) shows nonlinear conductance measurements of two devices. As earlier reported, both with rising density and conductance, the nonlinear conductance under a source-drain bias changes form, with both a single peak (type I ZBA) spontaneously split peaks (type II ZBA), clearly evident. The transitions between these peaks vary from sample to sample in seemingly random fashion, although all are restricted to fairly low densities. The dependence of ZBA on top gate bias was investigated, and though minimal effect was seen on the strength or form of the peaks, there was a change in the conductance value at which the peaks formed and split(Fig.\hspace{.1cm}\ref{Fig3}), implying that density ($k_F$) is causing transitions between ZBA type I and type II, and not value of conductance. The suppression of the ZBA with increasing temperature, as well as peak broadening in split peaks, (Fig.\hspace{.1cm}\ref{Fig4}(c,d)) is consistent with the proposed Kondo physics, as in linear splitting of single peaks in a magnetic field Fig.\hspace{.1cm}\ref{Fig4}(e).

In conclusion, we present a two level gating technique necessary to exploit the benefits  of a fully undoped heterostructure for mesoscopic devices. The effect of background impurities in large 2D regions is clearly demonstrated and quantified. We show that the length scale set by these unintentional impurities can have a signature in mesoscopic structures. The appearance of a zero bias anomaly is shown not to be critically dependent on the presence of intentional dopants. Future experiments will extend these processes and estimates to the investigation of ultra-clean 1D and 0D devices.

The authors acknowledge D. Anderson for electron beam patterning, and F. Sfigakis and A. Corcoles for useful discussions. This work was supported by EPSRC. S. Sarkozy acknowledges financial support as a Northrop Grumman Space Technology Doctoral Fellow and from the Cambridge Overseas Trust. C. Siegert acknowledges financial support from Gottlieb Daimler- and Karl Benz-Foundation.


\begin{thebibliography}{20}
\expandafter\ifx\csname natexlab\endcsname\relax\def\natexlab#1{#1}\fi
\expandafter\ifx\csname bibnamefont\endcsname\relax
  \def\bibnamefont#1{#1}\fi
\expandafter\ifx\csname bibfnamefont\endcsname\relax
  \def\bibfnamefont#1{#1}\fi
\expandafter\ifx\csname citenamefont\endcsname\relax
  \def\citenamefont#1{#1}\fi
\expandafter\ifx\csname url\endcsname\relax
  \def\url#1{\texttt{#1}}\fi
\expandafter\ifx\csname urlprefix\endcsname\relax\def\urlprefix{URL }\fi
\providecommand{\bibinfo}[2]{#2}
\providecommand{\eprint}[2][]{\url{#2}}
\bibitem[{\citenamefont{Pepper}(1981)}]{Pepper:Localization}
\bibinfo{author}{\bibfnamefont{M.}~\bibnamefont{Pepper}},
  \emph{\bibinfo{title}{Localization and interaction effects in a two
  dimensional electron gas}}, vol. \bibinfo{volume}{149}
  (\bibinfo{publisher}{Springer Berlin / Heidelberg}, \bibinfo{year}{1981}).

\bibitem[{\citenamefont{Imada et~al.}(1998)\citenamefont{Imada, Fujimori, and
  Tokura}}]{RevModPhys.70.1039}
\bibinfo{author}{\bibfnamefont{M.}~\bibnamefont{Imada}},
  \bibinfo{author}{\bibfnamefont{A.}~\bibnamefont{Fujimori}}, \bibnamefont{and}
  \bibinfo{author}{\bibfnamefont{Y.}~\bibnamefont{Tokura}},
  \bibinfo{journal}{Rev. Mod. Phys.} \textbf{\bibinfo{volume}{70}},
  \bibinfo{pages}{1039} (\bibinfo{year}{1998}).

\bibitem[{\citenamefont{Ando et~al.}(1982)\citenamefont{Ando, Fowler, and
  Stern}}]{2DScatteringRevModPhys.54.437}
\bibinfo{author}{\bibfnamefont{T.}~\bibnamefont{Ando}},
  \bibinfo{author}{\bibfnamefont{A.~B.} \bibnamefont{Fowler}},
  \bibnamefont{and} \bibinfo{author}{\bibfnamefont{F.}~\bibnamefont{Stern}},
  \bibinfo{journal}{Rev. Mod. Phys.} \textbf{\bibinfo{volume}{54}},
  \bibinfo{pages}{437} (\bibinfo{year}{1982}).

\bibitem[{\citenamefont{Gold}(1988)}]{GoldPhysRevB.38.10798}
\bibinfo{author}{\bibfnamefont{A.}~\bibnamefont{Gold}}, \bibinfo{journal}{Phys.
  Rev. B} \textbf{\bibinfo{volume}{38}}, \bibinfo{pages}{10798}
  (\bibinfo{year}{1988}).

\bibitem[{\citenamefont{Ando et~al.}(1977)\citenamefont{Ando, Fowler, and
  Stern}}]{AndoIR}
\bibinfo{author}{\bibfnamefont{T.}~\bibnamefont{Ando}},
  \bibinfo{author}{\bibfnamefont{A.~B.} \bibnamefont{Fowler}},
  \bibnamefont{and} \bibinfo{author}{\bibfnamefont{F.}~\bibnamefont{Stern}},
  \bibinfo{journal}{J. Phys. Soc. Japan} \textbf{\bibinfo{volume}{43}},
  \bibinfo{pages}{1616} (\bibinfo{year}{1977}).

\bibitem[{\citenamefont{Lilly et~al.}(2003)\citenamefont{Lilly, Reno, Simmons,
  Spielman, Eisenstein, Pfeiffer, West, Hwang, and
  Das~Sarma}}]{PhysRevLett.90.056806}
\bibinfo{author}{\bibfnamefont{M.~P.} \bibnamefont{Lilly}},
  \bibinfo{author}{\bibfnamefont{J.~L.} \bibnamefont{Reno}},
  \bibinfo{author}{\bibfnamefont{J.~A.} \bibnamefont{Simmons}},
  \bibinfo{author}{\bibfnamefont{I.~B.} \bibnamefont{Spielman}},
  \bibinfo{author}{\bibfnamefont{J.~P.} \bibnamefont{Eisenstein}},
  \bibinfo{author}{\bibfnamefont{L.~N.} \bibnamefont{Pfeiffer}},
  \bibinfo{author}{\bibfnamefont{K.~W.} \bibnamefont{West}},
  \bibinfo{author}{\bibfnamefont{E.~H.} \bibnamefont{Hwang}}, \bibnamefont{and}
  \bibinfo{author}{\bibfnamefont{S.}~\bibnamefont{Das~Sarma}},
  \bibinfo{journal}{Phys. Rev. Lett.} \textbf{\bibinfo{volume}{90}},
  \bibinfo{pages}{056806} (\bibinfo{year}{2003}).

\bibitem[{\citenamefont{Harrell et~al.}(1999)\citenamefont{Harrell, Pyshkin,
  Simmons, Ritchie, Ford, Jones, and Pepper}}]{harrell:2328}
\bibinfo{author}{\bibfnamefont{R.~H.} \bibnamefont{Harrell}},
  \bibinfo{author}{\bibfnamefont{K.~S.} \bibnamefont{Pyshkin}},
  \bibinfo{author}{\bibfnamefont{M.~Y.} \bibnamefont{Simmons}},
  \bibinfo{author}{\bibfnamefont{D.~A.} \bibnamefont{Ritchie}},
  \bibinfo{author}{\bibfnamefont{C.~J.~B.} \bibnamefont{Ford}},
  \bibinfo{author}{\bibfnamefont{G.~A.~C.} \bibnamefont{Jones}},
  \bibnamefont{and} \bibinfo{author}{\bibfnamefont{M.}~\bibnamefont{Pepper}},
  \bibinfo{journal}{Appl. Phys. Lett.} \textbf{\bibinfo{volume}{74}},
  \bibinfo{pages}{2328} (\bibinfo{year}{1999}).

\bibitem[{\citenamefont{Kane et~al.}(1993)\citenamefont{Kane, Pfeiffer, West,
  and Harnett}}]{kane:2132}
\bibinfo{author}{\bibfnamefont{B.~E.} \bibnamefont{Kane}},
  \bibinfo{author}{\bibfnamefont{L.~N.} \bibnamefont{Pfeiffer}},
  \bibinfo{author}{\bibfnamefont{K.~W.} \bibnamefont{West}}, \bibnamefont{and}
  \bibinfo{author}{\bibfnamefont{C.~K.} \bibnamefont{Harnett}},
  \bibinfo{journal}{Appl. Phys. Lett.} \textbf{\bibinfo{volume}{63}},
  \bibinfo{pages}{2132} (\bibinfo{year}{1993}).

\bibitem[{\citenamefont{Willett et~al.}(2006)\citenamefont{Willett, Pfeiffer,
  and West}}]{willett:242107}
\bibinfo{author}{\bibfnamefont{R.~L.} \bibnamefont{Willett}},
  \bibinfo{author}{\bibfnamefont{L.~N.} \bibnamefont{Pfeiffer}},
  \bibnamefont{and} \bibinfo{author}{\bibfnamefont{K.~W.} \bibnamefont{West}},
  \bibinfo{journal}{Appl. Phys. Lett.} \textbf{\bibinfo{volume}{89}},
  \bibinfo{eid}{242107} (\bibinfo{year}{2006}).

\bibitem[{\citenamefont{Lu et~al.}(2007)\citenamefont{Lu, Luhman, Lai, Tsui,
  Pfeiffer, and West}}]{lu:112113}
\bibinfo{author}{\bibfnamefont{T.~M.} \bibnamefont{Lu}},
  \bibinfo{author}{\bibfnamefont{D.~R.} \bibnamefont{Luhman}},
  \bibinfo{author}{\bibfnamefont{K.}~\bibnamefont{Lai}},
  \bibinfo{author}{\bibfnamefont{D.~C.} \bibnamefont{Tsui}},
  \bibinfo{author}{\bibfnamefont{L.~N.} \bibnamefont{Pfeiffer}},
  \bibnamefont{and} \bibinfo{author}{\bibfnamefont{K.~W.} \bibnamefont{West}},
  \bibinfo{journal}{Appl. Phys. Lett.} \textbf{\bibinfo{volume}{90}},
  \bibinfo{eid}{112113} (\bibinfo{year}{2007}).

\bibitem[{\citenamefont{Pfeiffer and West}(2003)}]{Pfeiffer:Growth}
\bibinfo{author}{\bibfnamefont{L.}~\bibnamefont{Pfeiffer}} \bibnamefont{and}
  \bibinfo{author}{\bibfnamefont{K.~W.} \bibnamefont{West}},
  \bibinfo{journal}{Physica E} \textbf{\bibinfo{volume}{20}},
  \bibinfo{pages}{57} (\bibinfo{year}{2003}).

\bibitem[{\citenamefont{Hwang and Sarma}(2008)}]{hwang:235437}
\bibinfo{author}{\bibfnamefont{E.~H.} \bibnamefont{Hwang}} \bibnamefont{and}
  \bibinfo{author}{\bibfnamefont{S.~D.} \bibnamefont{Sarma}},
  \bibinfo{journal}{Phys. Rev. B} \textbf{\bibinfo{volume}{77}},
  \bibinfo{eid}{235437} (\bibinfo{year}{2008}).

\bibitem[{\citenamefont{Mani et~al.}(2002)\citenamefont{Mani, Smet, {von
  Klitzing}, Narayanamurti, Johnson, and Umanskyk}}]{ManiNature}
\bibinfo{author}{\bibfnamefont{R.~G.} \bibnamefont{Mani}},
  \bibinfo{author}{\bibfnamefont{J.~H.} \bibnamefont{Smet}},
  \bibinfo{author}{\bibfnamefont{K.}~\bibnamefont{{von Klitzing}}},
  \bibinfo{author}{\bibfnamefont{V.}~\bibnamefont{Narayanamurti}},
  \bibinfo{author}{\bibfnamefont{W.~B.} \bibnamefont{Johnson}},
  \bibnamefont{and} \bibinfo{author}{\bibfnamefont{V.}~\bibnamefont{Umanskyk}},
  \bibinfo{journal}{Nature} \textbf{\bibinfo{volume}{420}},
  \bibinfo{pages}{646} (\bibinfo{year}{2002}).

\bibitem[{\citenamefont{Sarkozy et~al.}(2007)\citenamefont{Sarkozy, {Das
  Gupta}, Sfigakis, Farrer, Beere, Harrell, Ritchie, and
  Jones}}]{Sarkozy:Electrochem}
\bibinfo{author}{\bibfnamefont{S.}~\bibnamefont{Sarkozy}},
  \bibinfo{author}{\bibfnamefont{K.}~\bibnamefont{{Das Gupta}}},
  \bibinfo{author}{\bibfnamefont{F.}~\bibnamefont{Sfigakis}},
  \bibinfo{author}{\bibfnamefont{I.}~\bibnamefont{Farrer}},
  \bibinfo{author}{\bibfnamefont{H.~E.} \bibnamefont{Beere}},
  \bibinfo{author}{\bibfnamefont{R.~H.} \bibnamefont{Harrell}},
  \bibinfo{author}{\bibfnamefont{D.~A.} \bibnamefont{Ritchie}},
  \bibnamefont{and} \bibinfo{author}{\bibfnamefont{G.~A.~C.}
  \bibnamefont{Jones}}, \bibinfo{journal}{ECS Trans.}
  \textbf{\bibinfo{volume}{11}}, \bibinfo{pages}{75} (\bibinfo{year}{2007}).

\bibitem[{\citenamefont{Cronenwett et~al.}(2002)\citenamefont{Cronenwett,
  Lynch, Goldhaber-Gordon, Kouwenhoven, Marcus, Hirose, Wingreen, and
  Umansky}}]{Cronenwett:ZBA}
\bibinfo{author}{\bibfnamefont{S.~M.} \bibnamefont{Cronenwett}},
  \bibinfo{author}{\bibfnamefont{H.~J.} \bibnamefont{Lynch}},
  \bibinfo{author}{\bibfnamefont{D.}~\bibnamefont{Goldhaber-Gordon}},
  \bibinfo{author}{\bibfnamefont{L.~P.} \bibnamefont{Kouwenhoven}},
  \bibinfo{author}{\bibfnamefont{C.~M.} \bibnamefont{Marcus}},
  \bibinfo{author}{\bibfnamefont{K.}~\bibnamefont{Hirose}},
  \bibinfo{author}{\bibfnamefont{N.~S.} \bibnamefont{Wingreen}},
  \bibnamefont{and} \bibinfo{author}{\bibfnamefont{V.}~\bibnamefont{Umansky}},
  \bibinfo{journal}{Phys. Rev. Lett.} \textbf{\bibinfo{volume}{88}},
  \bibinfo{pages}{226805} (\bibinfo{year}{2002}).

\bibitem[{\citenamefont{Siegert et~al.}(2007)\citenamefont{Siegert, Ghosh,
  Pepper, Farrer, and Ritchie}}]{Siegert:NaturePhysics}
\bibinfo{author}{\bibfnamefont{C.}~\bibnamefont{Siegert}},
  \bibinfo{author}{\bibfnamefont{A.}~\bibnamefont{Ghosh}},
  \bibinfo{author}{\bibfnamefont{M.}~\bibnamefont{Pepper}},
  \bibinfo{author}{\bibfnamefont{I.}~\bibnamefont{Farrer}}, \bibnamefont{and}
  \bibinfo{author}{\bibfnamefont{D.~A.} \bibnamefont{Ritchie}},
  \bibinfo{journal}{Nature Physics} \textbf{\bibinfo{volume}{3}},
  \bibinfo{pages}{315} (\bibinfo{year}{2007}).

\bibitem[{\citenamefont{Sfigakis et~al.}(2008)\citenamefont{Sfigakis, Ford,
  Pepper, Kataoka, Ritchie, and Simmons}}]{sfigakis:026807}
\bibinfo{author}{\bibfnamefont{F.}~\bibnamefont{Sfigakis}},
  \bibinfo{author}{\bibfnamefont{C.~J.~B.} \bibnamefont{Ford}},
  \bibinfo{author}{\bibfnamefont{M.}~\bibnamefont{Pepper}},
  \bibinfo{author}{\bibfnamefont{M.}~\bibnamefont{Kataoka}},
  \bibinfo{author}{\bibfnamefont{D.~A.} \bibnamefont{Ritchie}},
  \bibnamefont{and} \bibinfo{author}{\bibfnamefont{M.~Y.}
  \bibnamefont{Simmons}}, \bibinfo{journal}{Phys. Rev. Lett.}
  \textbf{\bibinfo{volume}{100}}, \bibinfo{pages}{026807}
  (\bibinfo{year}{2008}).

\bibitem[{\citenamefont{Cronenwett et~al.}(1998)\citenamefont{Cronenwett,
  Oosterkamp, and Kouwenhoven}}]{Cronenwett07241998}
\bibinfo{author}{\bibfnamefont{S.~M.} \bibnamefont{Cronenwett}},
  \bibinfo{author}{\bibfnamefont{T.~H.} \bibnamefont{Oosterkamp}},
  \bibnamefont{and} \bibinfo{author}{\bibfnamefont{L.~P.}
  \bibnamefont{Kouwenhoven}}, \bibinfo{journal}{Science}
  \textbf{\bibinfo{volume}{281}}, \bibinfo{pages}{540} (\bibinfo{year}{1998}).

\bibitem[{\citenamefont{Schuster et~al.}(1994)\citenamefont{Schuster, Ensslin,
  Wharam, K\"uhn, Kotthaus, B\"ohm, Klein, Tr\"ankle, and
  Weimann}}]{SchusterPhysRevB.49.8510}
\bibinfo{author}{\bibfnamefont{R.}~\bibnamefont{Schuster}},
  \bibinfo{author}{\bibfnamefont{K.}~\bibnamefont{Ensslin}},
  \bibinfo{author}{\bibfnamefont{D.}~\bibnamefont{Wharam}},
  \bibinfo{author}{\bibfnamefont{S.}~\bibnamefont{K\"uhn}},
  \bibinfo{author}{\bibfnamefont{J.~P.} \bibnamefont{Kotthaus}},
  \bibinfo{author}{\bibfnamefont{G.}~\bibnamefont{B\"ohm}},
  \bibinfo{author}{\bibfnamefont{W.}~\bibnamefont{Klein}},
  \bibinfo{author}{\bibfnamefont{G.}~\bibnamefont{Tr\"ankle}},
  \bibnamefont{and} \bibinfo{author}{\bibfnamefont{G.}~\bibnamefont{Weimann}},
  \bibinfo{journal}{Phys. Rev. B} \textbf{\bibinfo{volume}{49}},
  \bibinfo{pages}{8510} (\bibinfo{year}{1994}).

\bibitem[{\citenamefont{Weiss et~al.}(1993)\citenamefont{Weiss, Richter,
  Menschig, Bergmann, Schweizer, von Klitzing, and
  Weimann}}]{WeissPhysRevLett.70.4118}
\bibinfo{author}{\bibfnamefont{D.}~\bibnamefont{Weiss}},
  \bibinfo{author}{\bibfnamefont{K.}~\bibnamefont{Richter}},
  \bibinfo{author}{\bibfnamefont{A.}~\bibnamefont{Menschig}},
  \bibinfo{author}{\bibfnamefont{R.}~\bibnamefont{Bergmann}},
  \bibinfo{author}{\bibfnamefont{H.}~\bibnamefont{Schweizer}},
  \bibinfo{author}{\bibfnamefont{K.}~\bibnamefont{von Klitzing}},
  \bibnamefont{and} \bibinfo{author}{\bibfnamefont{G.}~\bibnamefont{Weimann}},
  \bibinfo{journal}{Phys. Rev. Lett.} \textbf{\bibinfo{volume}{70}},
  \bibinfo{pages}{4118} (\bibinfo{year}{1993}).

\end{thebibliography}

\end{document}